\documentclass[twocolumn,aps,amsmath,showpacs,pre]{revtex4}
\usepackage{graphicx}
\usepackage{amssymb}

\begin{document}
\title{Nature of the collapse transition in interacting self-avoiding trails}

\author{Tiago J. Oliveira
\footnote{On leave at Ames Laboratory - USDOE and Department of Physics \& Astronomy, Iowa State University, Ames, Iowa 50011, United States}}
\email{tiago@ufv.br}
\affiliation{Departamento de F\'isica, Universidade Federal de Vi\c cosa, 36570-900, Vi\c cosa, MG, Brazil}
\author{J\"urgen F. Stilck}
\email{jstilck@if.uff.br}
\affiliation{Instituto de F\'{\i}sica and National Institute of Science and Technology for Complex Systems, Universidade Federal Fluminense, Av. Litor\^anea s/n, 24210-346, Niter\'oi, RJ, Brazil}

\date{\today}

\begin{abstract}
We study the interacting self-avoiding trail (ISAT) model on a Bethe lattice of general coordination $q$ and on a Husimi lattice built with squares and coordination $q=4$. The exact grand-canonical solutions of the model are obtained, considering that up to $K$ monomers can be placed on a site and associating a weight $\omega_i$ for a $i$-fold visited site. Very rich phase diagrams are found with non-polymerized (NP), regular polymerized (P) and dense polymerized (DP) phases separated by lines (or surfaces) of continuous and discontinuous transitions. For Bethe lattice with $q=4$ and $K=2$, the collapse transition is identified with a bicritical point and the collapsed phase is associated to the dense polymerized phase (solid-like) instead of the regular polymerized phase (liquid-like). A similar result is found for the Husimi lattice, which may explain the difference between the collapse transition for ISAT's and for interacting self-avoiding walks on the square lattice. For $q=6$ and $K=3$ (studied on the Bethe lattice only), a more complex phase diagram is found, with two critical planes and two coexistence surfaces, separated by two tricritical and two critical end-point lines meeting at a multicritical point. The mapping of the phase diagrams in the canonical ensemble is discussed and compared with simulational results for regular lattices.
\end{abstract}

\pacs{61.41.+e,05.40.Fb,05.70.Fh}

\maketitle

\section{Introduction}
\label{intro}

When a polymer is placed in a solution, eventually, as the temperature (or the solvent quality) is lowered, it undergoes a transition from an extended (coil) to a collapsed (globule) configuration, at the so called $\theta$ temperature \cite{f66}. This transition is continuous and in the parameter space of a grand-canonical formalism, it may be identified as a tricritical point \cite{dg75,dg79}.

In the standard lattice model to study this collapse transition, the polymer is represented by a self-avoiding walk (SAW) - where no site and no bond are visited more than once - and the presence of the solvent is taken into account in an effective manner by introducing an attractive interaction between nearest neighbor (NN) monomers which are not consecutive in the walk (we use the terminology of monomers, placed on sites of the lattice, connected by bonds which are on lattice edges).  
The grand-canonical solution of this interacting self-avoiding walk (ISAW) model on hierarchical (Bethe and Husimi) lattices exhibits the expected tricritical point \cite{pablo90}. Exact results for this  model on regular two-dimensional lattices lead to the tricritical exponents $\nu_{\theta}=4/7$ \cite{ds87}, and it is believed that $\nu_{\theta}=1/2$ in 3D, corresponding to the classical (mean-field) value, since $d=3$ is the upper critical dimension for a tricritical point \cite{f66,dg79}. These exponents - in the so-called $\theta$-universality class - are also found when interactions between \textit{next-nearest} monomers are included in the system \cite{Lee,nathann}. Interestingly, if the site (monomer) interaction is replaced by a bond interaction (between bonds on opposite edges of elementary squares) on the square lattice, the phase diagram may change, due to the appearance of a stable dense polymerized phase for the model defined on a $q=4$ Husimi lattice built with squares \cite{sms96} and on the square lattice \cite{mos01}.

\begin{figure}[!t]
\centering
\includegraphics[width=8.5cm]{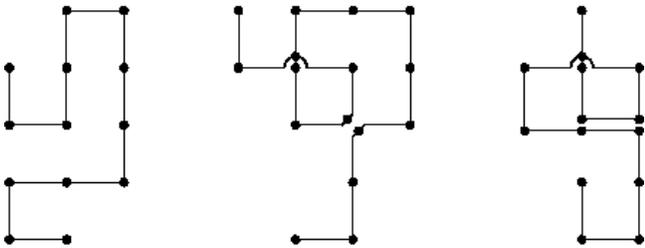}
\caption{Three walks on the square lattice. From left to right: a SAW, a SAT and a MMS configuration. The MMS walk and trail shown here visit each lattice site at most twice. For trails on the square lattice, this is the upper limit for the number of visits of a site.}
\label{Figwalks}
\end{figure}

An alternative to the two-site interactions of the ISAW model is to consider walks which may pass more than once through a lattice site, without any restriction on bonds, and associate a negative energy to multiply occupied sites. If up to $K$ monomers are allowed to occupy a site, we have the multiple monomer per site (MMS) model proposed by Krawczyk et al. \cite{kpor06}. Monte Carlo simulations of this model for $K=3$ on the square lattice suggest that it can display both continuous and discontinuous coil-globule transitions \cite{kpor06}. However, exact solutions of the MMS model on hierarchical lattices show that the collapse transition is always continuous, but its nature can be critical or tricritical, depending on the energies involved \cite{pablo07,tiago}. If the configurations of the MMS model are restricted such that each edge of the lattice is occupied at most by a single polymer bond, the resulting walks are called interacting self-avoiding trails (ISAT) \cite{moore}. Figure \ref{Figwalks} shows the differences among ISAW, ISAT and MMS chains. All SAW's are valid configurations of ISAT's, and all ISAT's are allowed in the MMS model, so that the models are in an order of increasing generalization. Notice that the maximum number of monomers per site $K$ in the MMS model can assume any value, while for ISAT this maximal number will be $q/2$ or $(q-1)/2$ for even or odd coordination number $q$, respectively. It is apparent in the figures that in all models the chains are linear, that is, each monomer which is not an endpoint is bonded to two other monomers. Nevertheless, the configurations of SAT resemble the ones found in branched polymers (BP), where a monomer may be bonded to more than two other monomers, which present rich phase diagrams \cite{dh83,bs95}. Indeed, as will be discussed below, on the Bethe lattice, BP with even ramification can be mapped on the ISAT model.

When a site is visited more than once in the ISAT model, there will be multiple ways to connect the incoming bonds, as may be seen in Fig. \ref{Figwalks}. If the trails are not allowed to cross themselves, we have the vertex-interacting self-avoiding walk (VISAW) model proposed by Bl\"ote and Nienhuis (BN) \cite{BN89}, whose coil-globule transition in two dimensions is associated to a different tricritical point, with $\nu_{BN}=12/23$ \cite{warnaar92}. This model presents a very rich phase diagram when the parameter space is increased by including stiffness \cite{vernier15}, so that the chains are semi-flexible. In contrast, for the more general ISAT model, where the trails are allowed to cross themselves, there are no exact results and the nature of its collapse transition is a subject of long debate in literature: while some works present evidences of continuous collapse transition in BN \cite{foster09} or ``undetermined" \cite{shapir84,Meirovitch,Lyklema,Owczarek} universality classes, the possibility of a discontinuous transition was also suggested in \cite{grassberger96}.
It seems to be no surprise that the inclusion of crossings in the BN model apparently makes it no longer exactly solvable and may lead to richer phase diagrams. For instance, if second neighbor (diagonal) interactions are introduced in the square lattice Ising model, since the lattice become non-planar, no exact solutions are known and a quite rich phase diagram is found, including a critical line with continuously varying critical exponents \cite{b79}.

In order to shed some light on these controversies, here we solve the ISAT model on Bethe and Husimi lattices considering that up to $K$ monomers can be placed on a site. Rich grand-canonical phase diagrams are found with non-polymerized, regular polymerized and dense polymerized phases separated by critical and coexistence lines/surfaces. Particularly for lattice coordination $q=4$ and $K=2$, which is an approximation for the ISAT on the square lattice, the solution on both lattices shows that the collapse transition is associated to a bicritical point. For the Bethe lattice with $q=6$ and $K=3$, which mimics cubic or triangular lattices, two tricritical and two critical lines meeting at a multicritical point are found. The mapping of the phase diagrams in the canonical ensemble is discussed, as well as their similarities with numerical results for regular lattices. Although of course solutions on hierarchical lattices are not suited to answer questions related to the universality class of phase transitions, since they lead to classical exponents, they may be useful to approximate the thermodynamic properties of the models, such as their phase diagrams and the nature of the phase transitions in the system. 

The rest of this work is organized as follows. In Sec. \ref{defmod}, we define the model in more detail and present its solution on the Bethe lattice in terms of recursion relations. The thermodynamic behavior of the model on this lattice is discussed in Sec. \ref{tpm}. The solution of the model on a four-coordinated Husimi lattice is presented in section \ref{Husimi}. Final discussions and conclusions may be found in Sec. \ref{conc}.

\section{Definition of the model and solution on the Bethe lattice in terms of recursion relations} 
\label{defmod}

\begin{figure}[!t]
\centering
\includegraphics[width=6cm]{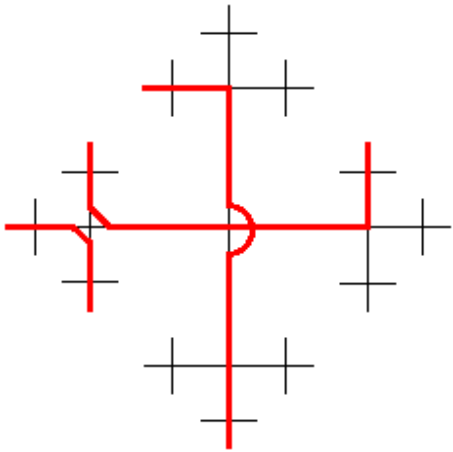}
\caption{(Color online) A contribution to the partition function of the model on a Cayley tree with $q = 4$ and 3 generations. The weight of this contribution will be $\omega_{1}^{9}\omega_{2}^{2}$. The endpoints of the walks are placed on the surface of the Cayley tree.} 
\label{Figrede}
\end{figure}

We consider interacting self-avoiding trails (ISAT's) on the Bethe lattice (the core of a Cayley tree \cite{baxter82}) with arbitrary coordination number $q$. In this model, at most one polymer bond can occupy a lattice edge. However, the lattice sites can be occupied by up to $K$ distinguishable monomers. A statistical weight $\omega_{n}$ is associated to each site with $n$ monomers. As usual, the endpoints of the walks are placed on the surface of the tree. The sites at the surface may be empty or have a monomer placed on them, and the statistical weights of these two configurations determine only the initial conditions of the solution of the model on treelike lattices in terms of recursion relations and, thus, they will have no influence on its thermodynamic behavior. The grand-canonical partition function of the model will be given by:
\begin{equation}
 Y = \sum_{N_{1},N_{2},\cdots,N_{K}} \omega_{1}^{N_{1}} \omega_{2}^{N_{2}} \cdots \omega_{K}^{N_{K}} 
\end{equation}
where the sum is over all configurations of the walks on the tree, while $N_{n}$ is the number of sites visited $n$ times by the walks. In Fig. \ref{Figrede}, an example of a Cayley tree with three generations of sites is shown, with a particular configuration of ISAT's which contributes to the partition function. As already noted, the maximum number of monomers that can be placed on a site is limited by the lattice coordination, being $K_{max} = q/2$ or $K_{max}=(q-1)/2$ for even and odd $q$'s, respectively.

To solve the model on the Bethe lattice we consider rooted subtrees, defining partial partition functions (ppf's) for them. A lattice edge may be empty or occupied by a trail, thus we define two ppf's $g_{0}$ and $g_1$ for empty and occupied root edges, respectively. Considering the operation of attaching $q-1$ subtrees with a certain number of generations to a new root site and edge, we build a subtree with an additional generation. The recursion relations for the two ppf's are:
\begin{subequations}
\begin{equation}
g'_{0}= g_{0}^{q-1} + \sum_{n=1}^{K} \binom{q-1}{2 n} \frac{(2 n-1)!}{2^{n-1}(n-1)!} \omega_{n} g_{0}^{q-2n-1} g_{1}^{2n},
\label{fpp0}
\end{equation}
\begin{equation}
g'_{1}= \sum_{n=1}^{K} \binom{q-1}{2 n-1} \frac{(2 n-1)!}{2^{n-1}(n-1)!} \omega_{n} g_{0}^{q-2n} g_{1}^{2n-1},
\label{fpp1}
\end{equation}
\end{subequations}
The first combinatorial factor in the expressions corresponds to the number of ways to choose the incoming bonds among the edges of the lattice, while the second term accounts for the number of ways to connect the incoming and the root bonds. These recursion relations, apart from the combinatorial factors, are identical to the ones obtained for branched polymers (BP) on the Bethe lattice \cite{bs95}, with the difference that the statistical weights (defined as $K_n$ there) are associated to sites with $n$ branches. As basically the thermodynamic properties of the model are obtained from the recursion relations, both models have similar behaviors, with some changes of variables (for even $n$'s). However, while Banchio and Serra \cite{bs95} restricted their analysis of the BP model for cases with two non-vanishing weights $K_m$ and $K_n$ only, here we will concentrate our attention mostly on the general case of the $K=3$ ISAT model with non-vanishing weights $\omega_1$, $\omega_2$ and $\omega_3$ for $q=6$, to compare with simulational results for ISAT's on the cubic and triangular lattices.

Usually, the partial partition functions diverge when the number $M$ of iterations (generations of the subtree) increases indefinitely. Thus, it is appropriate to define the ratio of them $R=g_{1}/g_{0}$, which is expected to remain finite in the thermodynamic limit ($M\rightarrow\infty$) at least for some range of the weight parameters ($\omega_i$) for which the density of empty sites on the lattice does not vanish. The recursion relation for the ratio is:
\begin{equation}
R' = \frac{\sum_{n=1}^{K} \binom{q-1}{2 n-1} \frac{(2 n-1)!}{2^{n-1}(n-1)!} \omega_{n} R^{2n-1}}{1 + \sum_{n=1}^{K} \binom{q-1}{2 n} \frac{(2 n-1)!}{2^{n-1}(n-1)!} \omega_{n} R^{2n}}.
\label{RR1}
\end{equation}
The properties of the model in the thermodynamic limit will be defined by a stable fixed point of the recursion relation $R'=R$. The fixed point is linearly stable if 
\begin{equation}
\frac{\partial R'}{\partial R} < 1.
\label{stab}
\end{equation}

The grand-canonical partition function of the model on the Cayley tree may be obtained if we consider the operation of attaching $q$ subtrees to the central site of the lattice, in a similar way as we used to derive the recursion relations for the ppf's. The result is:
\begin{equation}
Y=g_{0}^{q} y, \quad \text{with} \quad  y = 1 + \sum_{n=1}^{K} \binom{q}{2 n} \frac{(2 n-1)!}{2^{n-1}(n-1)!} \omega_{n} R^{2n}.
\label{gcpf1}
\end{equation}

The fraction of configurations with  $n$ monomers on the central site of the tree, which corresponds to the Bethe lattice result, is given by
\begin{equation}
 \rho_{n} = \frac{\omega_{n}}{Y} \frac{\partial Y}{\partial \omega_n} = \frac{\omega_{n}}{y} \binom{q}{2 n} \frac{(2 n-1)!}{2^{n-1}(n-1)!} R^{2n}
\label{densMono}
\end{equation}
and the total density of monomers on the lattice is
\begin{equation}
 \rho = \frac{1}{K}\sum_{n=1}^{K} n \rho_n.
\label{densTotal}
\end{equation}

The free energy of the model on the Bethe lattice is different from the one on the whole Cayley tree, since the contribution of the surface sites is discarded. Applying an ansatz proposed by Gujrati \cite{g95}, which may also be obtained considering the bulk and surface contributions to the free energy \cite{tiago}, the reduced free energy per site for the Bethe lattice is:
\begin{equation}
 \phi_{b} = -\frac{1}{2} \ln\left( \frac{Y_{M+1}}{Y_{M}^{(q-1)}} \right),
\end{equation}
in the limit of $M \rightarrow\infty$. Thus, from Eqs. (\ref{fpp0}) and (\ref{gcpf1}) we find
\begin{equation}
 \phi_{b} = -\frac{1}{2} \ln \left(  \frac{\left[ 1 + \sum_{n=1}^{K} \binom{q-1}{2 n} \frac{(2 n-1)!}{2^{n-1}(n-1)!} \omega_{n} R^{2n} \right]^{q}}{\left[ 1 + \sum_{n=1}^{K} \binom{q}{2 n} \frac{(2 n-1)!}{2^{n-1}(n-1)!} \omega_{n} R^{2n} \right]^{(q-2)}} \right).
\label{fe}
\end{equation}
It may be useful to recall that the grand-canonical potential $\Phi=-pV$, so that:
\begin{equation}
\phi_b=\frac{\Phi}{k_BTN}=-\frac{pv_0}{k_BT},
\label{fe1}
\end{equation}
where $N$ is the number of sites in the lattice and $v_0$ is the volume per site. 

\section{Thermodynamic properties of the model on the Bethe lattice}
\label{tpm}

It is easy to see in Eq. (\ref{RR1}) that $R=0$ is always a fixed point of the recursion relation. From Eqs. (\ref{densMono}) and (\ref{densTotal}), $\rho_i = 0$ and $\rho=0$ is obtained for $R=0$, so that this corresponds to a non-polymerized (NP) phase. Applying condition (\ref{stab}) to this fixed point, one see that it is stable for $\omega_1 \leqslant 1/(q-1)$. Furthermore, from Eq. (\ref{fe}), $\phi_b^{(NP)}=0$ is found.

Another possibility is that a fixed point is found for which $R \to \infty$. Expanding the recursion relations (\ref{RR1}) for large values of the ratio $R$, one notice that in general this fixed point will not be reached, since $R'$ is of the order of $R^{-1}$ in this limit. However, for trees with even values of $q$ and for $K=K_{max} = q/2$ the last term in the denominator of the recursion relation vanishes, and then
\begin{equation}
R' = \frac{\omega_{q/2}}{\omega_{q/2-1}}R+{\mathcal O}(R^{-1}),
\end{equation}
so that this fixed point will be stable if $\frac{\omega_{q/2}}{\omega_{q/2-1}} \ge 1$. Replacing the ratio $R \to \infty$ in the Eq. (\ref{densMono}), $\rho_n=\delta_{n,q/2}$ is found, so that all sites of the lattice are occupied by $q/2$ monomers in this dense polymerized (DP) phase and therefore all lattice edges are occupied by one bond. Actually, it is easy to obtain the free energy associated to this phase on any lattice, since every site will be occupied by $q/2$ monomers and all $qN/2$ edges of the lattice have bonds on them, so that it is necessary to consider only the number of ways to form pairs with the incoming bonds on each site. The result is:
\begin{equation}
\phi_b^{(DP)}=-\ln\left(\frac{(q-1)!}{2^{q/2-1}(q/2-1)!}\omega_{q/2}\right).
\label{fedp}
\end{equation}
On the Bethe lattice, this result may also be obtained writing a recursion relation for $S=1/R$ and studying the trivial fixed point $S=0$. The free energy Eq. (\ref{fe}) may be written in terms of $S$ and the result above for the free energy of the DP phase (Eq. (\ref{fedp})) is reobtained. Since $\phi_b^{(NP)}=0$, a NP-DP coexistence surface (where $\phi_b^{(DP)}=\phi_b^{(NP)}$) is expected to be located at
\begin{equation}
 \omega_{q/2} = \frac{2^{q/2-1}(q/2-1)!}{(q-1)!},
\label{coexNPDP}
\end{equation}
as indeed we will find below.

Finally, there may be other fixed points with non-vanishing and finite values of the ratio $R$, which may be found solving the polynomial fixed point equation for positive roots. This polymerized (P) phase in general corresponds to partial occupation of the lattice by the trails.

\subsection{Case of at most one monomer per site ($K=1$)}

For $K=1$, the SAW problem is recovered. This case has been studied before \cite{g84}, but we will briefly describe it here for completeness. Besides the NP fixed point, the recursion relation (\ref{RR1}) has the additional fixed point:
\begin{equation}
R^{P} = \sqrt{\frac{(q-1)\omega_{1} - 1}{\binom{q-1}{2} \omega_1}},
\end{equation}
which is associated to a polymerized phase. This phase is stable for $\omega_1 \geqslant 1/(q-1)$, so that a continuous polymerization transition occurs at the critical point $\omega_1 = 1/(q-1)$. Since for $q=3$ only $K=1$ is possible, this is the phase diagram of the ISAT model for this coordination number. For $q=3$, ISAT's are equivalent SAW's, as expected. It is also noteworthy that in the branched polymer (BP) model there exists the case of $3$ branches for $q=3$, which can not be mapped on ISAT model. We see that, even on Bethe lattice, these two models are not equivalent in general.

\subsection{Case of at most two monomers per site ($K=2$)}

When at most two monomers may occupy the same site, the fixed point equation for finite and non-vanishing values of the ratio $R$ is the biquadratic equation:
\begin{eqnarray}
3\binom{q-1}{4}\omega_2R^4 &+& 3\binom{q-1}{3}\left(\frac{\omega_1}{q-3}- \nonumber
\omega_2\right)R^2 \\
&+& 1-(q-1)\omega_1 =0.
\label{fpeK2}
\end{eqnarray}
For $q=4$, the first coefficient vanishes and the polymerized fixed point is
\begin{equation}
R^{P} = \sqrt{\frac{3 \omega_{1} - 1}{3(\omega_1-\omega_2)}},
\end{equation}
which admits real (physical) values only in the regions (\textit{I}) $\omega_1 \leqslant 1/3$ and $\omega_2 > \omega_1$, and (\textit{II}) $\omega_1 \geqslant 1/3$ and $\omega_2 < \omega_1$. The line $\omega_1=\omega_2$ is a stability limit of the phase P, since $J \equiv (\partial R'/\partial R)_{R^{P}}=1$ along it. In region \textit{I}, the P phase is unstable ($J>1$) and, thus, the P phase does not coexist with the NP phase. This is different from what is seen in the solution of the ISAW model on the Bethe lattice \cite{sms96}, where a NP-P coexistence line exists in this region. In region \textit{II} the P phase is stable ($J < 1$), with $J=1$ at $\omega_1 = 1/3$. Therefore, the line $\omega_1 = 1/3$ with $\omega_2 \leqslant \omega_1$ is a critical line separating the NP and P phases.

Recalling that the DP phase ($R \to \infty$) is stable for $\omega_2 \ge \omega_1$, we notice that the stability limit of the P phase coincides with the one of the DP phase at $\omega_2=\omega_1$, so that there is a continuous transition between these phases there (for $\omega_1 \geqslant 1/3$). When $\omega_1 \leqslant 1/3$ and $\omega_2 \geqslant \omega_1$, both NP and DP phases are stable, therefore these two phases coexist in this region. According to Eq. (\ref{coexNPDP}), for $q=4$, the coexistence line is given by $\omega_2=1/3$. 

\begin{figure}[!t]
\centering
\includegraphics[width=8cm]{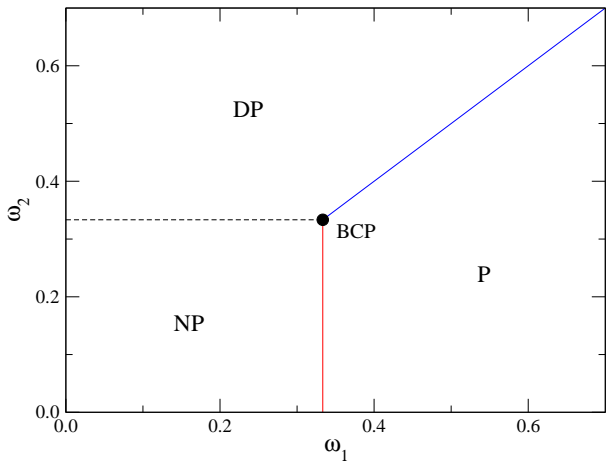}
\caption{(Color online) Phase diagram for $q=4$, with $K=2$. The full vertical (red) and slanted (blue) lines are the P-NP and P-DP critical lines, respectively. The NP and DP phases coexist on the dashed (black) line. The (black) circle is the bicritical point.} 
\label{figw1w2K2q4}
\end{figure}

These results are summarized in the phase diagram shown in Fig. \ref{figw1w2K2q4}, which is similar to the one for BP model in the case where the sites of the polymers are constrained to have only two or four incoming bonds \cite{bs95}. A bicritical point is located at $\omega_1=\omega_2=1/3$, where the two critical lines (NP-P) and (DP-P) meet the (NP-DP) coexistence line. The fact that the two critical lines meet at a finite angle indicates that the crossover exponent $\varphi$ associated to this bicritical point is equal to one, which is the classical value for this exponent, as expected. In contrast, in the ISAW model a tricritical ($\theta$) point is found in the phase diagram. Therefore, the fact that the collapse transition is of bicritical nature found here is in agreement with several works on square lattice showing that the universality classes of the collapse transition of the ISAT and the ISAW models are different \cite{foster09,Meirovitch,shapir84,Lyklema,Owczarek}.

For $q>4$, the first term in Eq. (\ref{fpeK2}) does not vanish, leading to a biquadratic equation, whose physical solutions are:
\begin{equation}
R_{\pm} = \left[  \frac{1}{6 b_4 \omega_2} \left( A \pm \sqrt{B}  \right) \right]^{1/2},
\end{equation}
where
\begin{equation} \nonumber
A \equiv 3 b_3 \omega_2 - b_2 \omega_1,
\end{equation}
\begin{equation} \nonumber
B \equiv A^2 + 12 b_4 \omega_2 \left[(q-1) \omega_1 - 1 \right],
\end{equation}
and the $b_i$'s are binomial coefficients
\begin{equation} \nonumber
b_i \equiv \binom{q-1}{i}.
\end{equation}
The first condition for these roots to be real is that $B>0$, which is satisfied for all $\omega_2$ if $\omega_1 \geqslant 1/(q-1)$. However, in this region $\sqrt{B}>A$, so that $R_{-}$ is complex. For $\omega_1 \leqslant 1/(q-1)$, both $R_{+}$ and $R_{-}$ admit physical solutions, but the last one is always unstable ($J\geqslant 1$, for all $\omega_2$ in this region). On the other hand, the solution $R_{+}$ for large enough $\omega_2$ is stable in this region and this polymerized (P) phase coexists with the NP one. Although we were not able to obtain a general expression for the limit of stability of the P phase in general, for a given $q$ it can be easily calculated. For instance, for $q=6$, we find $J \leqslant 1$ for $\omega_2 \geqslant 1/30 + \omega_1/6 + \sqrt {1 + 10 \omega_1 - 75 \omega_1^2}/30$ when $\omega_1 \leqslant 1/5$, and $J<1$ for all $\omega_2$ for $\omega_1 \geqslant 1/5$. At $\omega_1 = 1/5$, for $\omega_2 \leqslant 1/15$, we have $R_{+} = 0$ and $J=1$, so that the NP and P limits coincide and there exists a critical line ending at a tricritical point (TCP). This point can be obtained in general (for $q>4$). The result is:
\begin{equation}
\omega_1^{TCP} = \frac{1}{q-1}, \quad \quad \omega_2^{TCP} = \dfrac{1}{q^2-4 q+3}.
\end{equation}

\begin{figure}[!t]
\centering
\includegraphics[width=8cm]{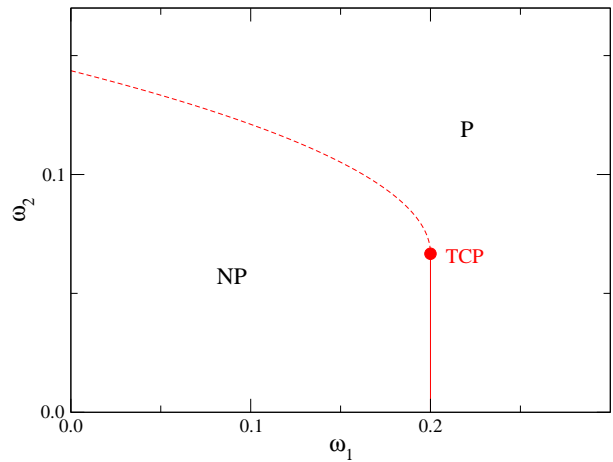}
\caption{(Color online) Phase diagram for $q=6$, with $K=2$. The full (vertical) and the dashed lines are the critical and coexistence lines, respectively. The circle is the tricritical point.}
\label{figw1w2K2q6}
\end{figure}

Above the TCP there exists a coexistence line. For $q=6$, the condition $\phi^{P}=\phi^{NP}=0$ leads to $\omega_2=1/15+(2/45) \sqrt{3-15 \omega_1}$. The resulting phase diagram for this coordination number is depicted in Fig. \ref{figw1w2K2q6} and similar ones are found for any $q>4$.

\subsection{Case of at most three monomers per site ($K=3$)}

In this case, we will solve the fixed point equation only for the case $q=6$, which is an approximation for ISAT's on triangular and cubic lattices. The finite and non-vanishing fixed point values of the ratio $R$ are the solutions of the biquadratic equation:
\begin{equation}
15 (\omega_2 - \omega_3) R^4 + 10 (\omega_1 - 3\omega_2) R^2 + 1-5\omega_1 = 0.
\label{fpe3}
\end{equation}
We will initially present some details of the general phase diagram by discussing slices at fixed values of $\omega_3$, $\omega_2$, and $\omega_1$.

\subsubsection{Phase diagrams for fixed $\omega_3$}

\begin{figure}[!t]
\centering
\includegraphics[width=8cm]{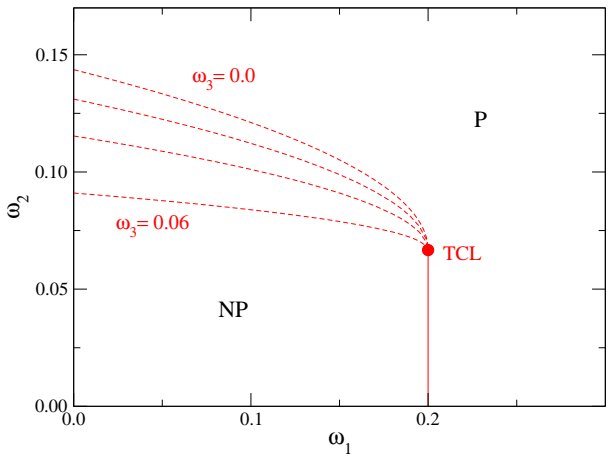}
\caption{(Color online) Phase diagrams for $q=6$, with $K=3$ and several values of $\omega_3<1/15$. The full (vertical) line represents the critical plane separating the NP and P phases. The circle indicate a tricritical line (TCL). The dashed lines represent the NP-P coexistence surface, from the top-to-bottom, lines for $\omega_3=0$, 0.02, 0.04 and 0.06 are shown.}
\label{figw1w2K3q6}
\end{figure}

Obviously, for $\omega_3=0$, the case $K=2$ above is recovered, so that the plane $\omega_1-\omega_2$ of the (three-dimensional) phase diagram is the one shown in Fig. \ref{figw1w2K2q6}. For $\omega_3>0$, it is not possible to obtain all critical and coexistence lines/surfaces analytically, but it is easy to determine them numerically. Figure \ref{figw1w2K3q6} shows phase diagrams for several values of $\omega_3<1/15$, which are qualitatively identical to the one obtained for $\omega_3=0$. Actually, only the NP-P coexistence lines change, forming thus a curved NP-P coexistence surface. Since the NP-P critical line stays at the same position, there exists a NP-P critical plane located at $\omega_1=1/5$, which meets the coexistence surface at a tricritical line (TCL) at $\omega_2 = 1/15$. Notice that, although the DP phase is stable for $\omega_3\geqslant \omega_2$, for $\omega_3<1/15$ the NP phase has a smaller free energy. 

As will be shown below, exactly at $\omega_3=1/15$ there exists a NP-DP coexistence plane limited by two critical-end-point lines, where the NP-P critical and coexistence surfaces end. Moreover, there is also a multicritical point on this plane where the TCL ends. 

Phase diagrams (not shown) for (fixed) $\omega_3>1/15$ present P-DP continuous and discontinuous transition lines meeting at a tricritical point (different from the NP-P one already discussed), so that there are also critical and coexistence P-DP surfaces as well as a second tricritical line in the 3D phase diagram. This will be demonstrated below.

\subsubsection{Phase diagrams for fixed $\omega_2$}

\begin{figure}[!t]
\centering
\includegraphics[width=8cm]{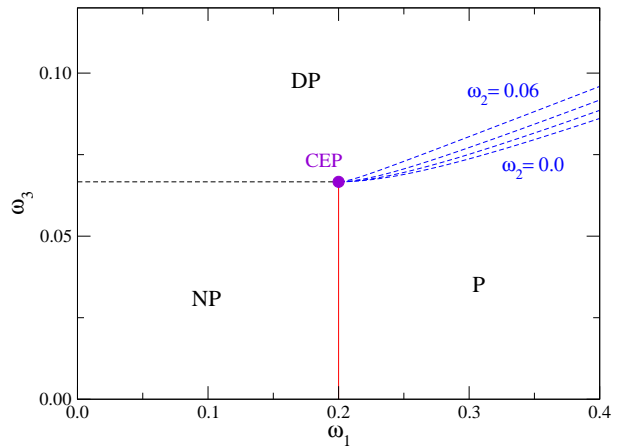}
\caption{(Color online) Phase diagrams for $q=6$, with $K=3$ and several $\omega_2<1/15$. The full vertical (red) line represents the critical P-NP plane. The dashed horizontal (black) line indicates the coexistence surface of NP and DP phases. The dashed (blue) lines are P-DP coexistence lines, from the bottom-to-top, for $\omega_2=0.0$, 0.02, 0.04 and 0.06. The (violet) circle represents the NP-P critical end-point line.}
\label{figw1w3K3q6}
\end{figure}

Considering $\omega_2=0$ in Eq. (\ref{fpe3}), besides the two fixed point values $R=0$ (NP) and $R \rightarrow \infty$ (DP), we have two other positive solutions of the fixed point equation:
\begin{equation}
R_{\pm} = \left( \frac{5\omega_1 \pm \sqrt{(5 \omega_1)^2 + 15\omega_3(1-5\omega_1)} }{15 \omega_3}    \right)^{1/2},
\end{equation}
but $R_{+}$ is unstable along the whole phase diagram.
In the region $\omega_1 < 1/5$, the solution $R_{-}$ is unphysical. At $\omega_1 = 1/5$, we have $R_{-}=0$ and $J(R_{-})=1$, leading to the expected NP-P critical line. For $\omega_1 > 1/5$, the solution $R_{-}$ (the P phase) is physical and stable for $\omega_3\leqslant 5 \omega_1^2 / 3(5\omega_1 -1)$. Notice that this stability limit diverges when $\omega_1 \rightarrow 1/5$ and, thus, the NP-P critical line, at $\omega_1 = 1/5$, exists for $\omega_3$ in the interval $[0,\infty)$. As discussed above, the DP phase is stable for $\omega_3 \geqslant \omega_2$, which in the present case means that this phase is stable in the whole $\omega_1-\omega_3$ plane, except at $\omega_3=0$. The NP-DP coexistence line (CL), given by Eq. (\ref{coexNPDP}), is located at $\omega_3=1/15$ (for $\omega_1\leqslant1/5$). The coexistence line between P-DP phases, obtained equating their free energies, is given by $\omega_3 = \left( 2-15\omega_1 + 2\sqrt{15\omega_1(5 \omega_1 - 1)+1} \right)/15$, for $\omega_1\geqslant 1/5$. This line meets the NP-DP coexistence line tangentially at $\omega_1=1/5$ and $\omega_3=1/15$, which is a critical end-point (CEP), where the NP-P critical line becomes metastable. These results are summarized in the phase diagram shown in Fig. \ref{figw1w3K3q6}. For $\omega_2>0$, the spinodals of the P phase (not shown in the figure) as well as the P-DP coexistence line were obtained numerically. In the region $\omega_2<1/15$ the same qualitative behavior seen for $\omega_2=0$ is found and only the P-DP coexistence line changes (see Fig. \ref{figw1w3K3q6}). Therefore, the curved P-DP and the plane NP-DP coexistence surfaces meet at a line of CEP located at $\omega_1=1/5$ and $\omega_3=1/15$. This line ends when it meets the tricritical one, at the multicritical point at $\omega_1=1/5$ and $\omega_2=\omega_3=1/15$.

\subsubsection{Phase diagrams for fixed $\omega_1$}

\begin{figure}[!t]
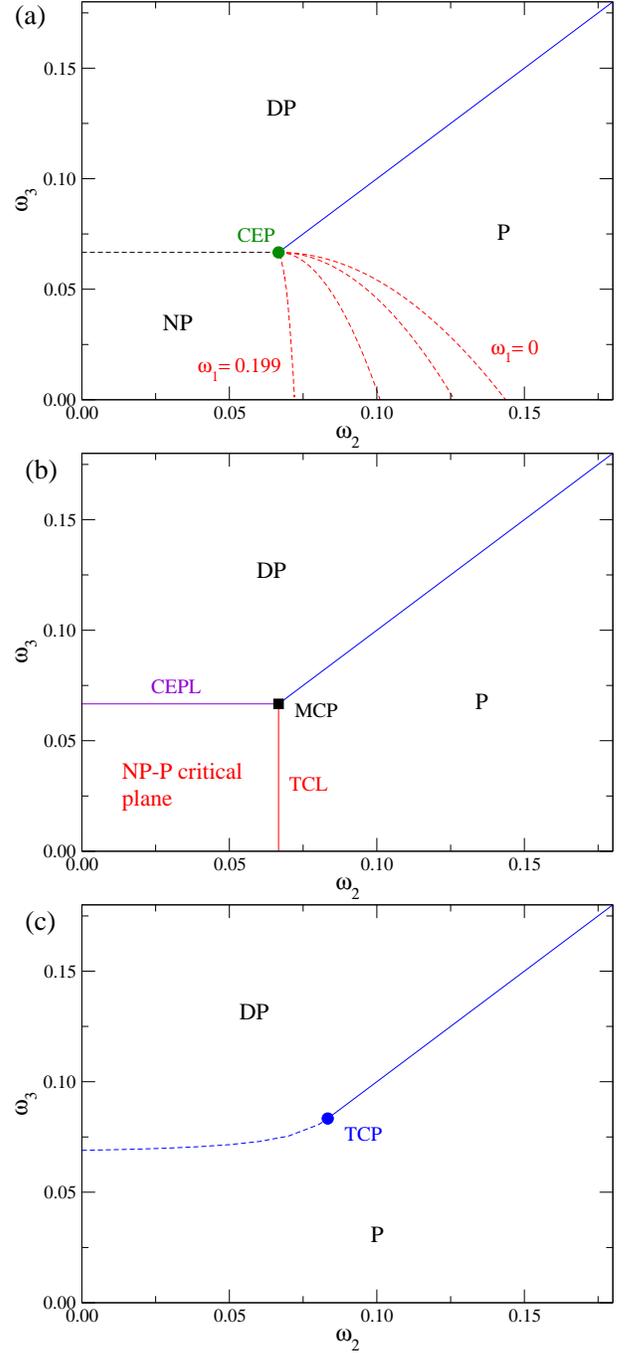

\centering
\includegraphics[width=7.99cm]{Diagq6K3w2w3.eps}
\includegraphics[width=7.99cm]{Diagq6K3w2w3w1020.eps}
\includegraphics[width=7.99cm]{Diagq6K3w2w3w1025.eps}
\caption{(Color online) Phase diagrams for $q=6$, with $K=3$ and (a) several values of $\omega_1<1/5$, (b) $\omega_1=1/5$ and (c) $\omega_1=0.25$. In (a) the full (slanted, blue) and dashed (horizontal, black) lines represent the critical P-DP and the coexistence NP-DP planes, respectively. The dashed (red) lines constitute the P-NP coexistence surface, from the right-to-left, lines for $\omega_1=0$, 0.08, 0.16 and 0.199 are shown. In (b), at the multicritical point (black square), the P-NP critical end-point line (horizontal, violet), the tricritical P-NP line (vertical, red) and the critical P-DP line (slanted, blue) meet. In (c), at the tricritical point (circle) the critical (solid) and coexistence (dashed) P-DP lines meet.}
\label{figw2w3K3q6}
\end{figure}

For $\omega_1=0$, Eq. (\ref{fpe3}) has the positive roots
\begin{equation}
R_{\pm} = \left( \frac{15\omega_2 \pm \sqrt{(15 \omega_2)^2 - 15 (\omega_2- \omega_3)} }{15 (\omega_2-\omega_3)}    \right)^{1/2},
\end{equation}
but the solution $R_{-}$ is unstable in the whole $\omega_2-\omega_3$ plane.
Notice that in the region \textit{I} ($\omega_3 < -15 \omega_2^2 + \omega_2$) both solutions are non-physical. It is easy to see that $R_{+}$ diverges when $\omega_2 \rightarrow \omega_3$ and, indeed, it is stable for $\omega_3 \leqslant \omega_2$ (except in region \textit{I}), corresponding to a polymerized (P) phase. This stability limit coincides with the one of DP phase, so that a DP-P critical line exists at $\omega_3 = \omega_2$. For $\omega_2 < 1/15$, the NP-DP phases coexist at $\omega_3=1/15$, while for $\omega_2 > 1/15$ the NP-P coexistence line is given by $\omega_3 = -45 \omega_2^2/4 + 3 \omega_2/2 +1/60$, which meets (tangentially) the NP-DP coexistence line, at $\omega_3=\omega_2=1/15$. At this point the P-DP critical line becomes metastable, so that it is also a critical end-point. The resulting phase diagram is shown in Fig. \ref{figw2w3K3q6}(a). For any $\omega_1$ in the range $[0,1/5)$, the same behavior is found, and only the NP-P coexistence surface changes, as expected (see Fig. \ref{figw2w3K3q6}(a)). Therefore, we find the expected NP-DP coexistence plane, at $\omega_3=1/15$ (for $\omega_2 \leqslant 1/15$). Moreover, the stability limits of the P and DP phases still meet at $\omega_3=\omega_2$, giving rise to a critical P-DP plane there, for $\omega_2\geqslant1/15$. This plane ends at a critical end-point (CEP) line, at $\omega_3=\omega_2=1/15$, where also the NP-P and NP-DP coexistence surfaces meet. This CEP line also ends at the multicritical point.

At the plane $\omega_1=1/5$, which is the stability limit of the NP phase, Eq. (\ref{fpe3}) has the polymerized solution
\begin{equation}
R = \sqrt{\frac{10 (3 \omega_2 - 1/5)}{15 (\omega_2 - \omega_3)}},
\end{equation}
which is stable for $\omega_3 \leqslant \omega_2$ and $\omega_2\geqslant 1/15$. The phase diagram at this plane is shown in Fig. \ref{figw2w3K3q6}b.

\begin{figure}[t]
\centering
\includegraphics[width=8cm]{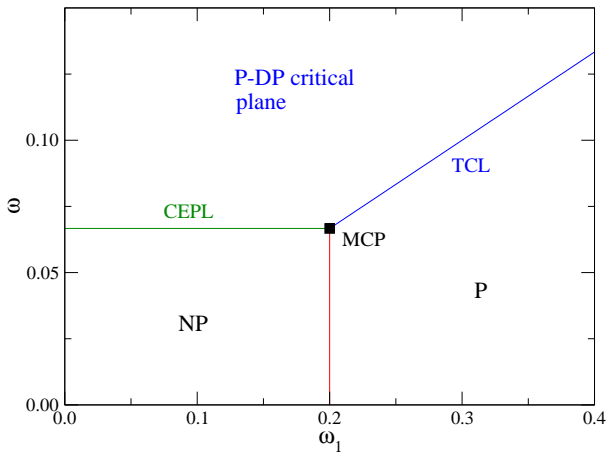}
\caption{(Color online) Phase diagram for $q=6$, with $K=3$ and $\omega=\omega_2=\omega_3$. At the multicritical point (black square) meet the P-DP critical end-point line (horizontal, green), the critical P-NP line (vertical, red) and the P-DP tricritical line (slanted, blue).}
\label{figw1wK3q6}
\end{figure}

For $\omega_1>1/5$, the NP phase becomes unstable and only P-DP transitions are found in the phase diagram. As expected from the results above, for small $\omega_2$ this transition is discontinuous, but it becomes continuous at a tricritical point, as shown in Fig. \ref{figw2w3K3q6}c (for $\omega_1=1/4$). Therefore, the P-DP critical and coexistence surfaces meet at a (second) tricritical line. 

\subsubsection{Phase diagram for $\omega_2 = \omega_3$}

Since the second tricritical line is located at the $\omega_3=\omega_2$ plane, we will find the phase diagram related to it. In this case, Eq. (\ref{fpe3}) reduces to a quadratic polynomial, whose polymerized solution is
\begin{equation}
R = \sqrt{\frac{5 \omega_1 - 1}{10 (\omega_1 - 3 \omega)}},
\end{equation}
where $\omega\equiv \omega_2=\omega_3$. As expected, for $\omega_1>3\omega$, $R=0$ when $\omega_1=1/5$ and $J(R)=1$. More important, in the region $\omega_1>1/5$, $R$ diverges and $J(R) \rightarrow 1$ when $\omega=\omega_1/3$, which is the tricritical line. This phase diagram is depicted in Fig. \ref{figw1wK3q6}.

\subsubsection{Three-dimensional phase diagram}

\begin{figure}[t]
\centering
\includegraphics[width=8.5cm]{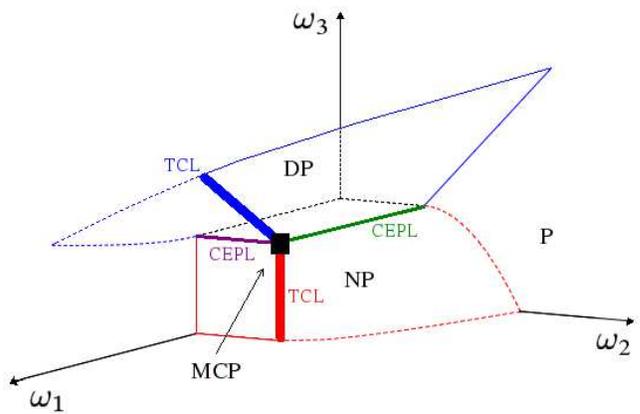}
\caption{(Color online) Sketch of the three dimensional phase diagram for $q=6$ with $K=3$. Colors and line type follow the same definitions from the previous figures.}
\label{figDiag3d}
\end{figure}

In summary, at the multicritical point the two tricritical lines and two critical end-point lines meet. The first ones separate critical and coexistence surfaces between P-NP and P-DP phases. These two critical surfaces are also limited by their corresponding CEP lines. The resulting three-dimensional phase diagram is depicted in Fig. \ref{figDiag3d}.

\section{Solution of ISAT on a Husimi lattice for $K=2$ and $q=4$}
\label{Husimi}

There are no closed paths on a Cayley tree, and therefore this is true also for the Bethe lattice. Thus, on these lattices, any random walk is a self avoiding walk. This may lead to results for the thermodynamic behavior which are qualitatively different from the ones found on regular lattices. 
A particular example, mentioned above, is the branched polymer model, which presents a bicritical behavior similar to the one shown in Fig. \ref{figw1w2K2q4} when considered on the Bethe lattice with $q=4$. However, its solution on a Husimi lattice built with squares (and $q=4$) yields a phase diagram where the bicritical point gives place to a critical-end-point and a tricritical point \cite{bs95}. It is therefore very important to obtain the thermodynamic behavior of the $K=2$ ISAT model on a $q=4$ Husimi lattice, to find out if the phase diagram also presents a similar change. We remark that, while the interchanging of crossing and colliding trails in the ISAT model on the Bethe lattice introduces only a coefficient in the recursion relation, on the Husimi lattice the existence of loops, even limited to squares, may forbid some configurations, breaking such equivalence between crosses and collisions.

\begin{figure}[!t]
\centering
\includegraphics[width=6cm]{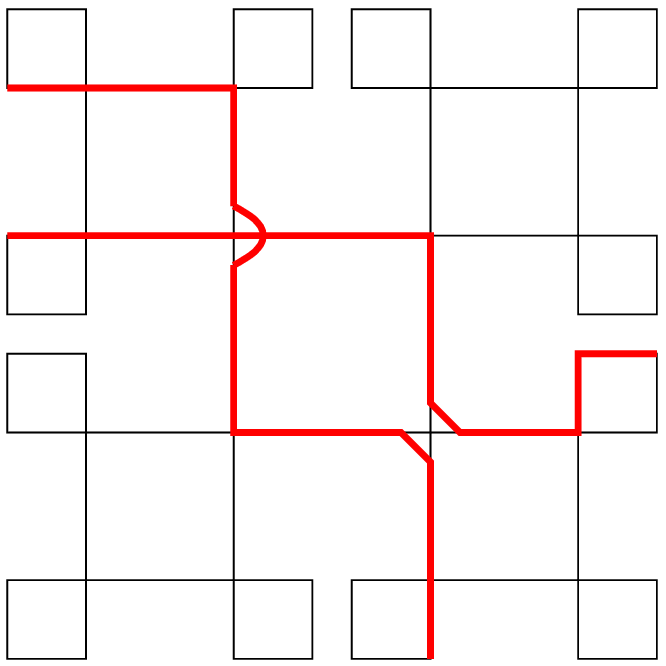}
\caption{(Color online) A contribution to the partition function of the model on a Husimi tree with $q = 4$ and 3 generations of squares. The weight of this contribution will be $\omega_{1}^{8}\omega_{2}^{2}$. The endpoints of the walks are placed on the surface of the tree.} 
\label{FigRedeH}
\end{figure}

\begin{figure}[!b]
\centering
\includegraphics[width=7.5cm]{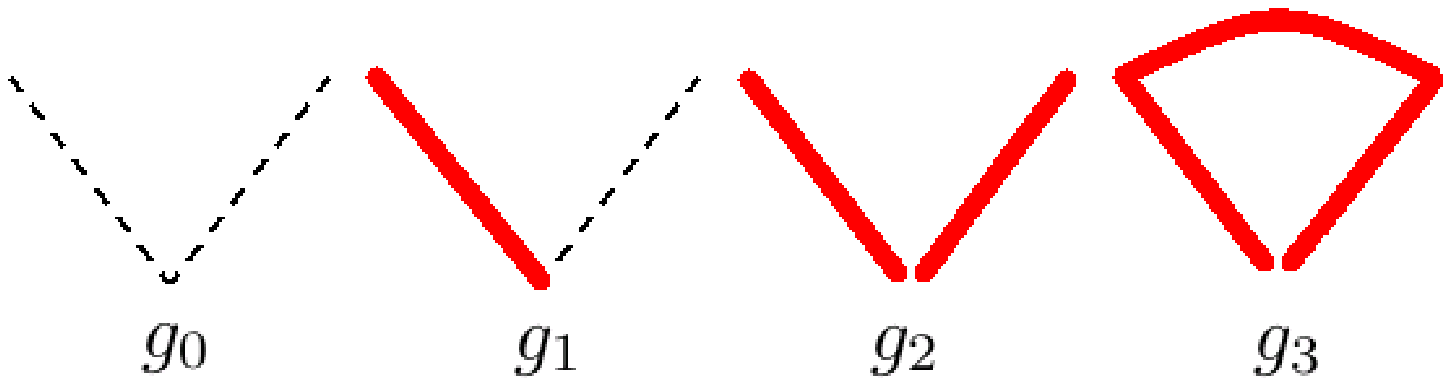}
\caption{(Color online) Definition of the root sites on the Husimi lattice. The thick (red) lines indicate incident bonds. The difference between $g_2$ and $g_3$ is that in the last (first) one the incident bonds are (not) connected.}
\label{FigppfsHusimi}
\end{figure}

The four coordinated Husimi tree, with three generations of squares, is shown in Fig. \ref{FigRedeH}, with two ISAT's placed on it. The solution of the model on the Husimi tree is done obtaining recursion relations for the partial partition functions on rooted subtrees, defined by the configurations of the site at their root. Four root configurations are needed, as shown in Fig. \ref{FigppfsHusimi}. We notice that it is necessary to distinguish between the ppf's 2 and 3: in the first the two bonds reaching the root site belong to different chains, starting at different sites of the surface of the tree, while in the second they are part of the same chain. This distinction is important to assure that no configuration with rings will be allowed in the model.

By considering the operation of attaching three subtrees to a new root square, the following recursions relation for the ppf's are found:

\begin{subequations}
\begin{eqnarray}
g'_{0} &=& a^3 + 2 a b^2 + b^2 c, \\
g'_{1} &=& 2 a^2 b + 2 a b c + 2 b c^2 + 2 b^3,\\
g'_{2} &=& a b^2 + 2 b^2 c + c^3 - d^3,\\
g'_{3} &=& d^3,
\end{eqnarray}
where
\begin{eqnarray}
a &=& g_0 + \omega_1 g_2,\\
b &=& \omega_1 g_1,\\
c &=& \omega_1 g_0 + \omega_2 (3 g_2 + 2 g_3),\\
d &=& \omega_1g_0+\omega_2(g_2+2g_3).
\end{eqnarray}
\label{Eqppfs}
\end{subequations}

Defining the ratios $R_{i}=g_{i}/g_{0}$, we obtain the following recursion relations for them:

\begin{subequations}
\begin{eqnarray}
R'_{1} &=& \frac{2 B ( A^2 + B^2 + C^2 + A C )}{E} ,\\
R'_{2} &=& \frac{A B^2 + 2 B^2 C + C^3 - D^3}{E},\\
R'_{3} &=& \frac{D^3}{E}.
\end{eqnarray}
\end{subequations}
where
\begin{equation}
E = A^3 + 2 A B^2 + B^2 C,
\end{equation}
and
\begin{subequations}
\begin{eqnarray}
A &=& 1 + \omega_1 R_2,\\
B &=& \omega_1 R_1,\\
C &=& \omega_1 + \omega_2 (3 R_2 + 2 R_3)\\
D &=& \omega_1+\omega_2(R_2+2R_3).
\end{eqnarray}
\label{EqRRs}
\end{subequations}

The partition function of the model on the Husimi tree is obtained considering the operation of attaching four subtrees to the central square. The result is:
\begin{equation}
Y=g_0^4[A^4+4B^2(A^2+AC+C^2)+2B^4+C^4-D^4].
\label{pfht}
\end{equation}
The densities of sites occupied by one or two monomers are given by:
\begin{subequations}
\begin{eqnarray}
\rho_1&=&\frac{\omega_1}{4 Y}\left(\frac{\partial Y}{\partial \omega_1}\right),\\
\rho_2&=&\frac{\omega_2}{4 Y}\left(\frac{\partial Y}{\partial \omega_2}\right),
\end{eqnarray}
\end{subequations}
where the ratios should have their fixed point values for the given statistical weights $\omega_1$ and $\omega_2$.

The free energy {\em per square} of the model on the Husimi lattice, which is the bulk of the Husimi tree, is again obtained using Gujrati's prescription described above:
\begin{eqnarray}
&\phi_b&=-\lim_{M \to \infty}\frac{1}{2} \ln \left(\frac{Y_{M+1}}{Y_M^3}\right)
\nonumber \\
&=&-\ln \left[ \frac{(A^3+2AB^2+B^2C)^2}{A^4+4B^2(A^2+AC+C^2)+2B^4+C^4-D^4} \right] .
\end{eqnarray}

\begin{figure}[t]
\centering
\includegraphics[width=8.5cm]{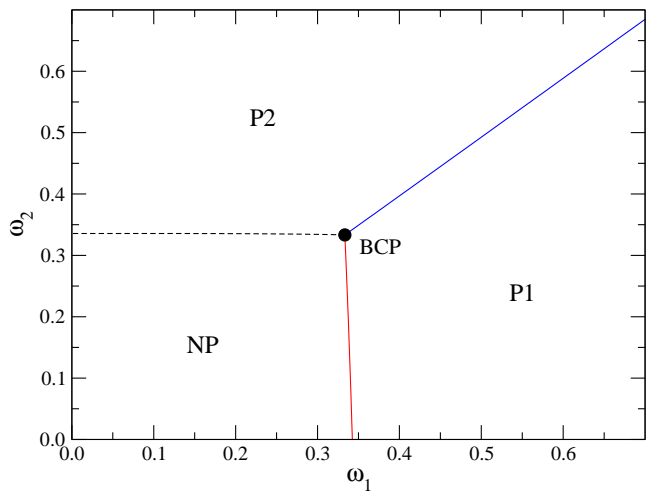}
\caption{(Color online) Phase diagram for the ISAT model on a Husimi lattice built with squares and coordination $q=4$. The full vertical (red) and slanted (blue) lines are the P1-NP and P1-P2 critical lines, respectively. The NP and P2 phases coexist on the dashed (black) line. The (black) circle is the bicritical point.}
\label{figDiagHusimi}
\end{figure}

Notice that these RR's can not be mapped on the ones for branched polymers \cite{bs95}, due to the existence of closed paths in the Husimi lattice. Despite this distinction, they have similar fixed points and thermodynamic phases:

\textit{i)} $R_1=R_2=0$ and $R_3 > 0$, which leads to $\rho_1=\rho_2=0$ and $\phi_b = 0$ and, thus, is a non-polymerized phase.

\textit{ii)} $R_1>0$, $R_2>0$ and $R_3>0$. In this polymerized phase all densities are non-vanishing. It will be called P1 here.

\textit{iii)} $R_1=0$, $R_2>0$ and $R_3 > 0$. This is also a polymerized phase, but more asymmetric than P1, typically with $\rho_1 \ll 1$ and $\rho_2 \approx 1$. This phase is similar to the dense one of Bethe lattice solution, however, it is not strictly dense, since, in general, $\rho_1 \neq 0$ and $\rho_2 \neq 1$. Therefore, we will refer to it as the P2 phase.

In contrast to the Bethe solution, here, all stability limits and coexistence lines have to be determined numerically. Continuous P1-NP and P1-P2 transition lines are found, while a coexistence line exists between the NP and P2 phases. All these lines meet at a bicritical point (BC) located at $\omega_1=\omega_2 \simeq 0.333333$. This point is probably $\omega_1=\omega_2=1/3$, exactly as in Bethe lattice, but we can not prove this analytically. These results are summarized in Fig. \ref{figDiagHusimi}, where the only difference observed when compared to the Bethe lattice diagram (Fig. \ref{figw1w2K2q4}) is that the transition lines are no longer linear.

\section{Final discussions and conclusions}
\label{conc}

The solution of the ISAT model on the Bethe lattice presented here confirms that the nature of collapse transition in this model can be, indeed, different from the one found for ISAW's (the $\theta$ class), depending on the interplay between the number of monomers allowed on a site ($K$) and the coordination $q$ of the lattice. For instance, for $K=K_{max}=q/2$, which is the case always considered on regular lattices, beyond a non-polymerized (NP) phase and a regular polymerized (P) phase, a stable dense polymerized (DP) phase may also exist. On the Bethe lattice, in this phase all sites of the lattice are occupied by $K_{max}$ monomers.

For $q=4$ and $K=2$, the P-NP critical line ends at a bicritical point, instead of the tricritical ($\theta$) point observed in the ISAW and MMS models. This bicritical behavior is different from all previous conjectures about the ISAT collapse transition on the square lattice, as discussed in the Introduction. It is important to recall that the same result has been found for the branched polymer (BP) model on the Bethe lattice, but on a Husimi lattice built with squares - which is a more realistic approximation to the square lattice - a different phase diagram was found, with the bicritical point given place to a critical endpoint and a tricritical point \cite{bs95}. In contrast, for the ISAT model on the Husimi lattice built with squares we found a phase diagram very similar to the one for the Bethe lattice (Fig. \ref{figw1w2K2q4}): the bicritical point is still present, and the only difference is the dense phase, which becomes a polymerized phase, but with asymmetric densities $\rho_1 \ll \rho_2$. This suggests that the collapse transition in the ISAT model on the square lattice may in fact be of bicritical nature.

It is also important to note that these results for ISAT's are very different from the ones for the MMS model (with $K=2$), where a tricritical point was found \cite{pablo07}. This shows that the restriction in the bonds is more important to determine the thermodynamic behavior of ISAT than the multiple visit of sites. Interestingly, for $K=2$ and $q>4$, the critical P-NP line ends at a tricritical point for the ISAT model, similarly to ISAW and MMS models. This may be compared with what happens for ISAW's with interactions between bonds: on the $q=4$ Husimi lattice the collapse transition is a critical endpoint, it becomes the usual tricritical point on a $q>4$ Husimi tree. We have, however, not studied the present model on Husimi trees with $q>4$.

Considering $q=6$ and $K=3$, a very rich phase diagram is found for Bethe lattice, with two tricritical (TC) lines separating the critical and coexistence P-NP and P-DP surfaces. Moreover, two critical end-point (CEP) lines are also present in the phase diagram, where the critical surfaces end. All these lines meet at a multicritical point (MCP) located at $\omega_1=1/5$, $\omega_2=\omega_3=1/15$. We recall that in the MMS model (with $K=3$) on the Bethe lattice there is also a critical NP-P surface limited by a TC line and a CEP line, both meeting at a MCP \cite{tiago}. In the version of the model where immediate reversal of the walk is forbidden the MCP is located at $\omega_i=1/(q-1)^i$ \cite{tiago}, differing from the location in ISAT. More important, in MMS model there is not a DP phase and the NP phase is limited by surfaces of continuous and discontinuous transitions to regular polymerized phases.

\begin{figure}[!t]
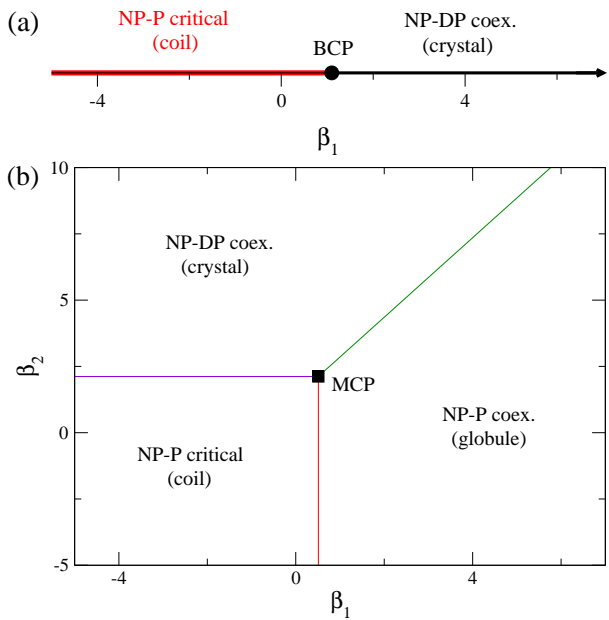

\centering
\includegraphics[width=8cm]{Diagq4K2canonico.eps}
\includegraphics[width=8cm]{Diagq6K3canonico.eps}
\caption{(Color online) Canonical phase diagrams for (a) $q=4$ with $K=2$ and (b) $q=6$ with $K=3$. In (b), the vertical (red), horizontal (violet) and slanted (green) lines are the P-NP TCL, P-NP CEPL and P-DP CEPL, respectively, which meet at the multicritical point (black square).}
\label{figDiagCanonico}
\end{figure}

In order to compare our results with simulations of ISAT on regular lattices, our grand-canonical phase diagrams can be mapped in canonical ones. As discussed in Ref. \cite{tiago}, in the canonical situation, the polymer chain is placed on an (effectively) infinite lattice, so the polymer (a polymerized phase) exists together with the empty lattice (a non-polymerized phase). Therefore, this corresponds to the surfaces limiting the NP phase in our diagrams. Moreover, identifying $\omega_1$ as the fugacity of the monomers, we must have $\omega_2=\omega_1^2 e^{\beta_1}$ and $\omega_3=\omega_1^3 e^{\beta_2}$, so that $\beta_1 = \ln\left[ \omega_2/\omega_1^2 \right] $ and $\beta_2 = \ln\left[ \omega_3/\omega_1^3 \right] $ are the canonical variables. Figure \ref{figDiagCanonico}(a) shows the canonical diagram for $q=4$ and $K=2$ and, as expected, increasing $\beta_1$ a collapse transition takes place at the bicritical point. However, for $\beta_1 > \beta_{1}^{*}$, instead of a globule (liquid-like) phase (due to a NP-P coexistence), we have found a dense (crystal-like) phase (due to a NP-DP coexistence). A dense phase has also been found in recent studies on generalized ISAT models on regular lattices \cite{doukas,foster11,bedini13} and is usually called a crystal phase. Thus, our results suggest that the origin of the difference in ISAW and ISAT models in the square lattice is the nature of their collapsed phases, which is liquid-like (globule) in the former and solid-like (crystal) in the last one. Noteworthy, the bicritical point is located at $\beta_{1}^{*} = \ln 3$ in striking agreement with the expected value for the collapse transition of the ISAT model on the square lattice \cite{Owczarek}.

The canonical diagram for $q=6$ and $K=3$ is depicted in Fig. \ref{figDiagCanonico}(b), where three phases are observed: coil, globule and crystal, all them separated by continuous transition lines that meet at the multicritical point (located at $\beta_1=\ln(5/3)$ and $\beta_2=\ln(25/3)$). Usually, equal energies are associated to sites visited twice and thrice, corresponding to the line  $\beta_1=\beta_2$ in our phase diagram. This line is placed inside the coil and globule phases only and, obviously, cuts the tricritical line at $\beta_1=\beta_2=\ln(5/3)$, suggesting that a transition similar to the ISAW model should be found in this case. 

In a very interesting work, Doukas et al. \cite{doukas} considered an extended ISAT model on the triangular lattice, where weights $\omega_2$ and $\omega_3$ were associated to double and triple visited sites. The canonical phase diagram they found, through Monte Carlo simulations (see Fig. 22 in Ref. \cite{doukas}), is qualitatively equal to the ours (Fig. \ref{figDiagCanonico}(b)), with the coil, globule and crystal phases and their respective transitions lines meeting at a multicritical point. The coil-globule transition was found to be continuous and belonging to the $\theta$ class, a continuous globule-crystal line is also found, in accordance with our findings. However, the coil-crystal transition is claimed to be first-order, while we have found a CEP line. Interestingly, a similar difference has been observed in the phase diagram of the MMS model, where a CEP line was found in the Bethe lattice solution \cite{tiago} and a first-order transition was suggested by Monte Carlo simulations \cite{kpor06}. Anyway, it is very interesting that comparing our results with the simulational findings by  Doukas et al. \cite{doukas}, we find that the locations for the multicritical point coincide. This quantitative agreement between solutions on hierarchical lattices and simulations is very rare and, as far as we known, it has been observed only in lattice gas systems \cite{pretti,tiagoGR}. Indeed, recent simulations of this generalized ISAT model on the cubic lattice showed that a dense phase does not exist in this case and, thus, a very different phase diagram is found \cite{bedini12}. This suggests that while our solution yields results reliable for the triangular lattice, it is not the case for the cubic one. Therefore, improved approximations, for example, solving the model on Husimi lattices built with cubes or triangles, are desirable to further study if this difference is found also on hierarchical lattices.

\section*{Acknowledgments}

We acknowledge the support of CNPq, FAPEMIG and FAPERJ (Brazilian agencies). TJO thanks the kind hospitality of the group of Prof. James Evans at Iowa State University, where part of this work was done.

\end{document}